%Paper: gr-qc/9403001
%From: Alan Rendall <anr@MPA-Garching.MPG.DE>
%Date: Tue, 1 Mar 1994 13:23:50 +0100 (MET)

\magnification=1200
\def\ip{\langle\ ,\ \rangle}
\def\R{{\bf R}}
\def\d{\partial}
\def\next{\hfil\break\noindent}
\font\title=cmbx12

{\title \centerline{
Adjointness relations as a criterion for choosing an inner product}}

\vskip .5 cm\noindent
Alan D. Rendall

\vskip .5cm\noindent
Max-Planck-Institut f\"ur Astrophysik, Karl-Schwarzschild-Str. 1,
Postfach 1523, 85740 Garching, Germany.

\vskip .5cm\noindent
{\bf 1. Sufficient conditions for uniqueness}

In the quantisation of constrained systems it can happen that one
obtains a representation of an algebra of quantum operators on a
vector space without a preferred inner product. Since an inner
product is necessary for the probabilistic interpretation of
quantum theory, some way needs to be found of introducing an
appropriate inner product on this vector space. If the classical
system being quantised possesses some background structure then
it may be possible to use this to fix an inner product. In the
case of gravity, where no background structure is present, this
is not an option. The algebra of quantum observables usually
admits a preferred *-operation, often related to complex conjugation
of functions on the classical phase space. It has been suggested
by Ashtekar that a preferred inner product could be fixed by the
requirement that this *-operation is mapped by the representation
into the operation of taking the adjoint of an operator with
respect to the inner product in question. Discussions of this
proposal can be found in [1-3]. The purpose of the following is
to discuss the circumstances under which this idea suffices to
determine the inner product uniquely.

Let $A$ be an associative algebra with identity over the complex
numbers. This is to be interpreted as the algebra of quantum
observables. Suppose that a representation $\rho$ of $A$ on a complex
vector space $V$ is given. Suppose further that a *-operation
$a\mapsto a^*$ is given on $A$. The defining properties of a
*-operation are that it is conjugate linear ($(\lambda a+\mu b)^*
=\bar\lambda a^*+\bar\mu b^*$), that $(ab)^*=b^*a^*$ and that
$(a^*)^*=a$. In this paper the origin of these various objects
will not be discussed; information on that can be found in
[1] and [2]. Instead we take this collection of objects as
starting point. The condition which is supposed to characterise
the inner product is that
$$\langle \rho(a)x,y\rangle=\langle x,\rho(a^*)y\rangle,\eqno(1)$$
for all $a\in A$ and $x,y\in V$. Note that there is a trivial
non-uniqueness due to the possibility of multiplying the inner
product by a non-zero real constant but this is not physically
significant since it leaves expectation values unchanged. In
order to ensure the uniqueness of the inner product it is
necessary to require that the representation $\rho$ be irreducible
in some appropriate sense. In [3] it was argued that the correct
concept of irreducibility to use is that of topological
irreducibility and this leads to the following definition:

\noindent
{\bf Definition 1} Let $A$ be a complex *-algebra with identity
and $\rho$ a representation of $A$ on a complex vector space
$V$. An inner product $\langle\ ,\ \rangle$ on $V$ is called
{\it strongly admissible} if

\noindent
(i) $\rho$ is a *-representation with respect to this inner
product i.e. equation (1) is satisfied

\noindent
(ii) for each $a\in A$ the operator $\rho(a)$ is bounded with
respect to the norm associated to the given inner product so
that $\rho$ extends uniquely by continuity to a representation
$\hat\rho$ on the Hilbert space completion $\hat V$ of $V$
with respect to this norm

\noindent
(iii) $\hat\rho$ is topologically irreducible i.e. it leaves
no non-trivial closed subspaces of $\hat V$ invariant

\vskip 10pt
This definition is tailored to the case of representations by
bounded operators; the unbounded case will be discussed later.
In [3] it was claimed that if $\langle\ ,\ \rangle_1$ and
$\langle\ ,\rangle_2$ are two inner products which are strongly
admissible for a given representation then
there exists a positive real number $c$ such that
$\langle\ ,\ \rangle_2=c\langle\ ,\ \rangle_1$. In fact this
is incorrect. An explicit example where the claim fails is
given in section 3 below. An additional condition which
ensures uniqueness will now be presented but first some
terminology is required. Let $\ip_1$ and $\ip_2$ be two
admissible inner products. The inner product $\ip_2$ will be
said to be {\it compatible} with $\ip_1$ if any sequence
$\{x_n\}$ in $V$ which satisfies $\langle x_n,x_n\rangle_1\to 0$
as $n\to\infty$ and which is a Cauchy sequence with respect to
$\ip_2$ also satisfies $\langle x_n,x_n\rangle_2\to 0$. (This
is not a standard definition. It is only introduced for
convenience in this paper.) The modified claim is now:

\noindent
{\bf Theorem 1} Let $\ip_1$ and $\ip_2$ be inner products on a
complex vector space $V$ which are strongly admissible with
respect to a representation $\rho$ of a complex *-algebra
$A$. Suppose that $\ip_2$ is compatible with $\ip_1$. Then
$\ip_2=c\ip_1$ for some positive real number $c$.

\vskip 10pt
This theorem will be proved in section 2. A question which
comes up immediately is: what is the interpretation of the
property of compatibility and is it a reasonable condition
from the point of view of the original motivation, namely
quantisation of certain systems? To make contact with known
mathematics, let $\hat V_1$ be the completion of $V$ with
respect to $\ip_1$ and consider $\ip_2$ as an unbounded
sesquilinear form on the Hilbert space $\hat V_1$ with domain
$V$. In general, if $S$ is an unbounded positive sesquilinear
form on a Hilbert space $H$ with domain $D$ then a new
inner product can be defined on $D$ by
$\langle x,y\rangle_S=\langle x,y\rangle_H+S(x,y)$.
The sesquilinear form $S$ is called closed if $D$ is complete
with respect to $\ip_S$. More generally $S$ is called closable
if it has an extension to a domain $\bar D$ such that the
extension is closed. It turns out that $\ip_2$ is compatible
with $\ip_1$ if and only if $\ip_2$ is closable when considered
as an unbounded sesquilinear form on $\hat V_1$ (see chapter 6
of [4]). It is useful
to have this information since it puts at our disposal known
results on closable sesquilinear forms.

Concerning the relation of the compatibility condition with the
original motivation, note first that the kind of situation
which is likely to occur in practice is that one inner product
is already known and we would like to know if it is characterised
by the condition that the *-relations go over to adjointness
relations. When the requirement of compatibility is added to
the other hypotheses of the theorem the result is not to restrict
the class of representations covered but rather to narrow the
class of alternative inner products within which uniqueness is
shown to hold. In examples it is usually the case that the
vector space $V$ is a space of functions on some set $X$ and that
the inner product which is known is the $L^2$ inner product
corresponding to some measure $d\mu$ on that set. It is desired to
have uniqueness within the class of inner products corresponding
to the $L^2$ norms defined by measures of the form $fd\mu$, where
$f$ is a non-negative real-valued function on $X$. There are general
theorems which make it reasonable to expect that these inner products
will all be compatible with the original one for conventional choices
of the space of functions $V$. Thus the compatibility condition does
not restrict the applicability of Theorem 1 too drastically.
To illustrate this some examples will now be discussed.

First let $(X,d\mu)$ be a general measure space and $f$ a non-negative
measurable real-valued function on $X$. Let
$$D_f=\left\{g\in L^2(X,d\mu):\int f|g|^2 d\mu <\infty\right\}.\eqno(2)$$
Then $D_f$ is dense in $L^2(X,d\mu)$ and the formula
$$S_f(g,h)=\int fg\bar h d\mu\eqno(3)$$
defines a closed sesquilinear form with domain $D_f$[4]. It is compatible
with the restriction of the $L^2$ norm to $D_f$ and so this provides
a wide class of examples where compatibility is satisfied. It is not
necessary to choose $V=D_f$ in this example. Any linear subspace
of $D_f$ which is dense in $L^2$ would also suffice. For instance in
the case that $X=\R^n$ and $d\mu$ is Lebesgue measure $V$ could be
chosen to be the space of smooth functions with compact support. More
generally, it is possible to consider two non-negative real-valued
locally integrable functions $f_1$ and $f_2$ on $\R^n$. Let $\ip_1$
and $\ip_2$ be the $L^2$ inner products defined by the measures
$f_1d\mu$ and $f_2d\mu$, restricted to $V=C_c^\infty(\R^n)$. A
sufficient condition that these sesquilinear forms
be inner products is that the zero sets of $f_1$ and $f_2$ have
zero measure. Assume that this is the case. $V$ is dense in
$L^2(\R^n,f_1d\mu)$ and it follows from the above that $\ip_2$
is compatible with $\ip_1$.

If one inner product is continuous with respect to the other,
$\|\ \|_2\le C\|\ \|_1$ for some constant $C$, then it is
obviously compatible with it. However it is in any case rather
easy to prove a uniqueness result for inner products in that
case. If $V$ is finite dimensional then the continuity is
automatic. It is natural to ask what happens when $V$ has a
countable basis. (The word basis is used here in the algebraic
and not in the Hilbert space sense.) Suppose that an inner product
$\ip_1$ is given on $V$. Using the Gram-Schmidt process it is possible
to go over to an orthonormal basis. Hence there is an isomorphism
of $V$ with the space of complex sequences with finitely many
non-zero entries such that the inner product takes the form
$\langle \{a_n\}, \{b_m\}\rangle_1=\sum a_n\bar b_n$. The completion
of $V$ can then be identified with the space $l^2$ of square summable
sequences and this will be done from now on. Any other inner product
on $V$ takes the form $\langle \{a_n\},\{b_m\}\rangle_2=\sum
a_nk_{mn}\bar b_m$ for some $k_{mn}$. Deciding whether a given set of
coefficients $k_{mn}$ defines an inner product compatible with
the original one is a concrete problem on the convergence of
sequences. Nevertheless it does not seem easy to give a general
solution. It will be shown by example in section 3 that the
compatibility condition does not always hold.

An analogue of Theorem 1 for unbounded operators will now be
presented. In [3] a procedure was given for reducing the
problem of uniqueness of the inner product in the case of a
representation by unbounded operators to the corresponding
problem for bounded operators under certain circumstances.
This reduction process combined with Theorem 1 gives a theorem
in the case of unbounded operators which will now be stated.

\noindent
{\bf Definition 2} Let $A$, $\rho$ and $V$ be as in Definition 1.
Let $S$ be a set of elements of $A$ which satisfy $a^*=a$ and
which generate $A$. An inner product $\ip$ on $V$ is said to
be {\it admissible} if:
\next
(i) $\rho$ is a *-representation with respect to this inner
product
\next
(ii) for each $a\in S$ the operator $\hat\rho(a)$ is essentially
self-adjoint
\next
(iii) $\hat\rho$ is irreducible
\next
(iv) $\hat\rho$ is closed

\vskip 10pt\noindent
Here $\hat\rho$ is obtained from $\rho$ by considering the linear
maps $\rho(a)$ on $V$ as unbounded operators on $\hat V$ with
domain $V$. The meanings of the words \lq closed\rq\  and
\lq irreducible\rq\ in this context are explained in [3]; suffice
it to say that this definition reduces to the definition of
\lq strongly admissible\rq\ in the case that all $\rho(a)$ are
bounded.

\noindent
{\bf Theorem 2} Let $\ip_1$ and $\ip_2$ be inner products on a
complex vector space $V$ which are admissible with
respect to a representation $\rho$ of a complex *-algebra
$A$. Suppose that $\ip_2$ is compatible with $\ip_1$. Then
$\ip_2=c\ip_1$ for some positive real number $c$.

\vskip .5cm\noindent
{\bf 2. Proof of the uniqueness theorem}

Let $\ip_1$ and $\ip_2$ be inner products satisfying the assumptions
of Theorem 1. Let $\ip=\ip_1+\ip_2$. Define $\hat V$ and $\hat V_1$
to be the completions of $V$ with respect to $\ip$ and $\ip_1$
respectively. The representation $\rho$ extends uniquely by
continuity to representations $\hat\rho$ and $\hat\rho_1$ on
$\hat V$ and $\hat V_1$ respectively. Now some facts proved in
[3] will be recalled. It was shown there under the hypotheses
that the inner products $\ip_1$ and $\ip_2$ are admissible
that there exists a bounded self-adjoint operator $L_1$ on
$\hat V$ such that $\langle x,y\rangle_1=\langle x,L_1y\rangle$
for all $x$, $y$ in $V$. It was also shown that unless the
two inner products are proportional the operator $L_1$ has
a non-trivial kernel. Now the sesquilinear form $\ip_1$ on $V$
is bounded with respect to $\ip$. Hence it extends uniquely by
continuity to a sesquilinear form $S$ on $\hat V$. Using
continuity again shows that $S(x,y)=\langle x,L_1y\rangle$ for
all $x$ and $y$ in $\hat V$. It follows that any vector $x$ in
the kernel of $L_1$ satisfies $S(x,x)=0$ and hence that $S$ is
degenerate. In [3] it was claimed that this is incompatible
with the fact that $\ip_1$ is an inner product. However this
is not true, as can be seen explicitly in the example given
in section 3 below. The facts that $\ip_1$ is non-degenerate
and that $S$ is an extension by continuity of $\ip_1$ do not
together imply that $S$ is non-degenerate.

Suppose then that $x\in\hat V$ satisfies the condition that
$S(x,x)=0$. Since $V$ is dense in $\hat V$ there exists a
sequence $x_n$ of vectors in $V$ with $\|x-x_n\|\to 0$ as
$n\to \infty$. On the other hand, the continuity of $S$ implies
that $S(x_n,x_n)\to 0$ and, due to the fact that all $x_n$ belong
to $V$, this is equivalent to the condition that $\|x_n\|_1\to 0$.
The sequence $x_n$ is a Cauchy sequence with respect to $\ip$
and hence with respect to $\ip_2$. Now the hypothesis that
$\ip_2$ is compatible with $\ip_1$ implies that $x_n\to 0$
with respect to $\ip_2$. We already know that it tends to
zero with respect to $\ip_1$. It follows that $x=0$. This
means that the kernel of $L_1$ is trivial and, in conjunction
with what was said above, completes the proof of the theorem.

\vskip .5cm\noindent
{\bf 3. A cautionary example}

Let $H$ be the Hilbert space $L^2([-1,1])$ and denote its inner
product by $\ip$. Let $V$ be the space of functions on $[-1,1]$
which extend analytically to a neighbourhood of that interval.
If $f\in V$ let $M_f$ be the multiplication operator on $H$
defined by $g\mapsto fg$. This is a bounded operator. If $\phi$
is an orientation preserving diffeomorphism of $[-1,1]$ with
$\phi(0)=0$ which extends to an analytic mapping on a neighbourhood
of $[-1,1]$ define an operator $T_\phi$ on $H$ by $T_\phi f=f\circ\phi$.
This operator is also bounded. Define $A$ to be the algebra of bounded
operators on $H$ generated by all the $M_f$ and $T_\phi$. If $\phi$
is a diffeomorphism as above define a function $\tilde\phi$ by
$$\tilde\phi(x)=\bar{d/dx(\phi^{-1}(x))}\eqno(4)$$
Then the adjoints of the operators of interest are given by $M_f^*=
M_{\bar f}$ and $T_\phi^*=M_{\tilde\phi}T_{\phi^{-1}}$. It follows
that the operation of taking the adjoint defines a *-operation on $A$.
Each operator belonging to $A$ maps $V$ into itself and so restricting
the elements of $A$ to $V$ defines a representation of $A$ on $V$. It
will be shown that there exist two inner products on $V$ which are
strongly admissible with respect to $\rho$ and which are not
proportional. These are defined as follows.
$$\eqalign{
\langle f,g\rangle_+=\int_0^1 f(x)\bar g(x) dx,         \cr
\langle f,g\rangle_-=\int_{-1}^0 f(x)\bar g(x) dx.}$$
These expressions obviously define sesquilinear forms but it needs
to be checked that they are non-degenerate on $V$. If $\langle f,f
\rangle_+=0$ then $f$ vanishes almost everywhere on $[0,1]$. But
by analyticity this implies that $f$ vanishes identically. The
proof for $\ip_-$ is similar.

It remains to show that both inner products are admissible. Because
of the symmetry of the situation it suffices to do this for $\ip_+$.
First note that the completion of $V$ with respect to $\ip_+$ can be
identified with $L^2([0,1])$. The computations which show that $\rho$
is a *-representation with respect to $\ip_+$ and that the operators
$M_f$ and $T_\phi$ are bounded with respect to the corresponding norm
$\|\ \|_+$ are essentially the same as those which are needed to show
that they are bounded on $H$ and to compute their adjoints there.
Next the irreducibility of $\hat\rho_+$ will be examined. Let $\Pi$
be a projection in the Hilbert space $\hat V_+$ which commutes with
all operators in the image of $\hat\rho_+$. The aim is to show that
$\Pi$ must be zero or the identity. Let $p=\Pi(1)$. This is an $L^2$
function on $[0,1]$. If $f$ belongs to $V$ then
$$pf=M_fp=M_f\Pi(1)=\Pi M_f(1)=\Pi f\eqno(5)$$
Thus on $V$ the operator $\Pi$ is given by multiplication by $p$.
It can easily be seen by approximating an arbitrary continuous
function on $[0,1]$ uniformly by elements of $V$ that in fact
$\Pi f=pf$ for any continuous function $f$. It will now be shown
that the function $p$ must be essentially bounded. Let $E_n$ be
the set where $|p|\ge n$ and let $\epsilon_n$ be the measure of
$E_n$. By Lusin's theorem [5] there exists a continuous function
$f_n$ with $|f_n|\le 1$ such that the measure of the set where
$f_n$ is not equal to the characteristic function of $E_n$ is
less than $\epsilon_n/2$. Hence $\|pf_n\|^2_2\ge n^2\epsilon_n /2$. On
the other hand $\|pf_n\|_2^2=\|\Pi f_n\|_2^2\le\|f_n\|_2^2\le
3\epsilon_n/2$. Hence $\epsilon_n=0$ for $n\ge 2$ and $p$ is
essentially bounded. It follows that multiplication by $p$ defines
a bounded operator $M_p$ on $L^2([0,1])$. Since this operator
agrees with $\Pi$ on a dense subspace it follows that $\Pi=M_p$.
Now $\Pi^2=\Pi$ implies that $M_p=M_p^2 =M_{p^2}$. Hence $(p^2-p)f=0$
for all $f\in L^2([0,1])$. It follows that $p=0$ or $1$ almost everywhere
and that $p$ is equal to the characteristic function $\chi_E$ of some
measurable subset $E$ of $[0,1]$. The condition $\Pi T_\phi=T_\phi\Pi$
will now be used. When worked out explicitly it gives
$$[p(x)-p(\phi(x))]f(\phi(x))=0\eqno(6)$$
for any $L^2$ function $f$. It follows that $p(\phi(x))=p(x)$ i.e.
that $\phi(E)=E$ up to set of measure zero. It will now be shown that
if $E$ has non-zero measure then it must differ from $[0,1]$ by a
set of zero measure. If $E$ has non-zero measure there must exist
a point $x\in E\cap(0,1)$ which is a point of density. A point of
density is roughly speaking a point of $E$ which is almost entirely
surrounded by other points of $E$; the exact definition can be found
in [6] where it is proved that any set of non-zero measure contains
such a point. This notion is invariant under diffeomorphisms and is
insensitive to altering the set $E$ by a set of measure zero. We
can therefore conclude that the set of points of density of
$E\cap (0,1)$ is invariant under all diffeomorphisms of the type
under consideration here. Consider now the vector field on $\R$
given by $X=(1-\cos(2\pi x))\d/\d x$. Exponentiating it gives a
one-parameter group of analytic diffeomorphisms $\phi_t$. The
restriction of each $\phi_t$ to the interval $[-1,1]$ belongs to
the class of diffeomorphisms used in defining the operators $T_\phi$.
Moreover, if $x_1$ and $x_2$ are any two points of $(0,1)$ there
is some $t$ for which $\phi_t(x_1)=x_2$. It follows that all points
of $(0,1)$ are points of density of $E$. The definition of points
then implies that no point of $(0,1)$ is a point of density
of the complement of $E$. Hence the complement of $E$ has measure
zero. This completes the proof of the irreducibility of $\hat\rho_+$
and hence of the strong admissibility of $\ip_+$.

A small modification of this example can be used to establish
another point, namely the existence of inner products on a
vector space with a countable basis which are not compatible.
Let $V^\prime$ be the vector space of polynomials on $[-1,1]$.
Define two inner products $\ip_+$ and $\ip_-$ on $V^\prime$ as
the restrictions of the corresponding inner products on $V$
defined above. Then $\ip_-$ is not compatible with $\ip_+$.
To see this, let $f$ be a continuous function on $[-1,1]$
which vanishes identically on $[0,1]$ but not on $[-1,0]$.
By Weierstrass' theorem there exists a sequence of polynomials
converging uniformly to $f$ on the interval $[-1,1]$. This
converges to zero with respect to the norm $\|\ \|_+$ and
is Cauchy with respect to $\|\ \|_-$. However it does not
converge to zero with respect to $\|\ \|_-$. The operator
algebra defined above does not act on $V^\prime$ and this
example is only meant to illustrate the notion of compatibility
of inner products. It is not known to the author whether Theorem 1
remains true for a vector space $V$ with countable basis if the
compatibility hypothesis is dropped.

\vskip 10pt\noindent
{\it Acknowledgements}

\noindent
I thank Horst Beyer for helpful remarks.

\vskip .5cm\noindent
{\bf References}
\next
1.Ashtekar, A. (1991) Lectures on non-perturbative canonical gravity.
World Scientific, Singapore.
\next
2.Ashtekar, A., Tate, R.S. An extension of the Dirac program for the
quantisation of constrained systems. Preprint.
\next
3.Rendall, A. D. (1993) Unique determination of an inner product by
adjointness relations in the algebra of quantum observables. Class.
Quantum Grav. {\bf 10}, 2261-2269.
\next
4.Kato, T. (1966) Perturbation Theory for Linear Operators. Springer,
Berlin.
\next
5.Rudin, W. (1987) Real and complex analysis. 3rd Edition.
McGraw-Hill, New York.
\next
6.Stein, E. M. (1970) Singular integrals and differentiability
properties of functions. Princeton University Press, Princeton.

\end